\documentclass[twocolumn,nofootinbib,prd,aps,showpacs,preprintnumbers,tightenlines]{revtex4}
\usepackage{epsfig} 
\begin{document}
\title{Neutral pion decay into $\nu\bar{\nu}$
in dense skyrmion matter}
\author{Alexander C. Kalloniatis$^a$}
\email{akalloni@physics.adelaide.edu.au}
\author{Jonathan D. Carroll$^a$}
\email{jcarroll@physics.adelaide.edu.au}
\author{Byung-Yoon Park$^{a,b}$}
\email{bypark@cnu.ac.kr}
\affiliation{$^a$ CSSM, University of
Adelaide, Adelaide 5005, Australia
 \\
$^b$ Department of Physics,Chungnam National University, 
Daejon 305-764, Korea }
\date{\today}
\preprint{ADP-05-01/T611}
\begin{abstract}
We study the weak decay of the neutral pion to a neutrino-antineutrino 
pair, $\pi^0\rightarrow \nu \bar{\nu}$, in the Skyrme model. 
In baryon free-space the process is forbidden by helicity 
while in a dense baryonic medium, the process 
becomes possible already to leading order in $G_F$
due to the break-down of Lorentz symmetry in the background 
medium. 
\end{abstract}

\maketitle

\section{Introduction}
Neutrinos and their means of production
play an important role in cosmological and
astrophysical phenomena \cite{Astrophysics}. 
In particular in this paper we shall be
concerned with the weak emission of a neutrino-antineutrino pair
from a neutral pion, $\pi^0\rightarrow \nu \bar{\nu}$, 
which figures centrally in the pion pole mechanism \cite{FGRVSHK76,FGRVSHK77} 
for the process  
$\gamma\gamma\rightarrow \pi^0 \rightarrow \nu \bar{\nu}$ 
relevant to the neutrino emission  
process $\gamma\gamma\rightarrow\nu\bar{\nu}$.
The latter has been suggested 
as a competitor to the modified URCA process for neutron
star cooling for most densities 
and for temperatures around $10^9$ K \cite{Po59,CM60} while it 
has been shown to 
be forbidden to lowest order in $G_F$ by Gell-Mann \cite{Ge61}.   
On the other hand, medium modifications of the pion can also
influence the modified URCA processes, as shown in \cite{VS86}, 
where neutrino-pair radiation from nucleons in 
$\pi^0$-condensed nucleon matter was studied. 
However, \cite{Lei04} has argued that such modifications
may not be so efficient
because of the gap between the energy bands of nucleons
interacting with
the periodic structure of the condensed pion field.

Returning then to the pion-pole mechanism, of course 
in the limit of zero neutrino mass the amplitude for the 
corresponding process vanishes 
for local weak currents of the $V-A$ form.
In \cite{FGRVSHK77} it was shown that a deviation from vanishing
neutrino mass could lead to small neutrino emission 
rates suppressed by a factor $(m_\nu/m_\pi)^3$, which however 
can be relevant in astrophysical processes for temperatures where 
the pion pole is significant. 
In \cite{LNg91}, the pion pole mechanism is used to 
determine a bound on the branching ratio of $R < 2.9 \times 10^{-13}$ 
for the decay $\pi^0 \rightarrow \nu\bar{\nu}$,
as well as an upper bound on the neutrino mass as $m_\nu < 420$ keV.
In \cite{ANV03}, various medium effects in the pion-pole 
mechanism $\gamma\gamma \rightarrow \pi^0 \rightarrow \nu\bar{\nu}$ 
have been discussed. 
In particular, the nuclear medium effects on the pion
propagator are estimated to determine the nuclear matter to vacuum
{\it ratio} of energy loss from pion-pole neutrino emission, such
that the decay width $\Gamma(\pi^0 \rightarrow \nu\bar{\nu})$
is cancelled but assumed to be non-zero either
from a neutrino mass or some other mechanism. 
The enhancement seen in this ratio of energy
loss rates for densities twice that of nuclear matter is of the order
of $10^4$ coming from the pion propagator modifications
in dense nuclear matter alone. In this paper we shall consider
precisely the baryonic medium modifications of $\pi^0 \rightarrow \nu\bar{\nu}$,
adding to the picture established thus far by \cite{ANV03}.

The simple reason that $\pi^0 \rightarrow \nu\bar{\nu}$ is forbidden 
in the limit of massless neutrino is the helicity selection rule. 
Two outgoing neutrinos have opposite chirality, which combine to total 
angular momentum one in the rest frame of the decaying meson. 
Thus, it is impossible for the pseudoscalar pion to decay into 
$\nu\bar{\nu}$. However in a dense medium such a 
process has been shown to be possible \cite{JPS02,RST03,VS86}. 
Of course, a scalar meson at rest with respect to the dense medium 
cannot decay into $\nu\bar{\nu}$ pairs. However,
a background medium provides an explicit 
reference frame so a scalar meson moving with respect to the medium 
can have higher angular momentum admixtures in its wavefunction. 
In \cite{JPS02} neutrino emissivity is studied
in the colour-flavour locked (CFL) phase at extreme
densities via decay of the ``generalised pion''
which is the lightest Goldstone boson in this phase. 
The non-vanishing decay comes from a breakdown of Lorentz symmetry 
which is explicitly incorporated into an effective Lagrangian via
differing temporal and spatial components of the pion decay constant, 
$f_T$ and $f_S=f_T v_\pi^2$ where $v_{\pi}^2$
represents an ``in-medium pion velocity''.
The resulting neutrino pair emissivity from $\pi^0$ decay
does not vanish and is proportional to $(f_T-f_S)^2$.

In this work, we study then the decay of the neutral pion to
$\nu\bar{\nu}$ specifically in a dense baryonic medium,
namely we work at densities below the critical point
for restoration of chiral symmetry.
As in previous works \cite{LPMRV03,LPRV03,LPRV04,PRV04,KP04}, 
we describe the baryonic matter via
the Skyrme model where both pions and dense baryonic matter 
are described by a single effective chiral Lagrangian. 
The basic strategy of the approach begins with the Skyrme  
conjecture \cite{Sk61} that a soliton (skyrmion) 
of the meson Lagrangian can be taken as a baryon 
so that dense baryonic matter can be approximated as 
a system of infinitely many skyrmions. 
Pions can be incorporated either as fluctuations on
the baryon free vacuum or on a classical field
configuration describing dense matter.
The dynamics of the pions become then strongly dependent 
on the background properties such as the baryon density. 
If we accept with \cite{SkyrmionCrystal} that the Skyrme model 
can be applied up to 
some density, its unique feature of a unified meson-baryon 
description provides an elegant framework for investigating
nonperturbative meson properties in dense baryonic matter.
In particular, we need not assume any density dependence of 
the in-medium parameters but rather work with a single model Lagrangian 
whose parameters are fixed for mesons in free space. 
Only the classical ground state solution describing the dense skyrmion matter 
becomes density dependent as do naturally in turn
the fluctuating mesons on top of this dense baryonic background.

The weak vector bosons are incorporated into the model Lagrangian 
by standard gauging. In baryon free space the 
$\pi^0-Z$ vertex from the kinetic term of the Lagrangian leads to a  
trivially vanishing amplitude for the process 
$\pi^0 \rightarrow Z \rightarrow \nu\bar{\nu}$. 
A new vertex for $\pi^0\rightarrow ZZ$ arises from the gauging of the 
WZW term. This also yields a vanishing result for
$\pi^0 \rightarrow ZZ \rightarrow \nu\bar{\nu}$ however
at the level of the amplitude-squared, which we shall show in this paper.
Most significantly we show in this paper
that for neutral pion fluctuations in  
dense skyrmion matter the dynamics of the pions are 
affected by the background and the $\pi^0 \rightarrow \nu\bar{\nu}$ 
can occur already to order $G_F$ with
a strength determined by nonperturbative dynamics to be of 
order $10^{-3}$, significantly larger than the $(m_{\nu}/m_{\pi})^3$
behaviour for keV neutrinos. 
As in \cite{JPS02,RST03}, the nonvanishing amplitude is ascribed to 
pion dynamics in the presence of the  
breakdown of Lorentz symmetry however now in the {\it hadronic}
rather than the CFL phase.
Unlike \cite{VS86} which also works in the hadronic phase,
in the present work the nuclear medium will modify the
pion, closer in spirit to \cite{ANV03}
where nevertheless these influences on $\pi^0 \rightarrow \nu\bar{\nu}$
were not considered in the hadronic phase. 

A difference between our approach and that of \cite{JPS02,RST03}
is that our initial Lagrangian itself respects Lorentz invariance;
the breaking of the symmetry rather takes place at the level of the
background field {\it ansatz} over which the pions fluctuate.
In view of the assumptions behind the skyrmionic model for
baryonic matter, our results are qualitative but are
summarised in Fig.4 where enhancement as a function of density
of the decay $\pi^0 \rightarrow \nu\bar{\nu}$ is indicated
in the quantity $\delta$, which is also intimately related
to the in-medium velocity of the pion $v_{\pi}$.
We compare these specific results to those of \cite{JPS02}
as well as to our previous work on $v_{\pi}$ in \cite{LPRV04}. 

This paper is organised as follows: in the next section
we introduce the model Lagrangian, followed by the
demonstration that $\pi^0\rightarrow \nu {\bar \nu}$ vanishes
in free space. In section IV we analyse the process in dense
skyrmion matter and give a brief statement of conclusions
at the end.

\section{Model Lagrangian}
We work with the Skyrme model Lagrangian \cite{Sk61}
\begin{eqnarray}
  {\cal L} &=& 
    \frac{f_\pi^2}{4}
    \mbox{Tr} (\partial_\mu U^\dagger \partial^\mu U) 
    + {\cal L}_{\mbox{\scriptsize sk}}
  \nonumber\\
  &&
    +\frac{f_\pi^2 m_\pi^2}{4} 
    \mbox{Tr} (U^\dagger + U - 2) 
    + {\cal L}_{\mbox{\scriptsize WZW}}  
\label{L0}
\end{eqnarray}
where $U=\exp(i\vec{\tau}\cdot\vec{\pi}/f_\pi) \in SU(2)$, 
$m_\pi$ and $f_\pi$ are 
respectively the pion mass and decay constant in free space. 
The Skyrme term ${\cal L}_{\mbox{\scriptsize sk}}$ 
is a higher derivative term introduced into the Lagrangian 
to stabilise the soliton solution, namely
\begin{equation}
  {\cal L}_{\mbox{\scriptsize sk}} 
  = \frac{1}{32e^2} \mbox{Tr}
  \left( [ L_\mu, L_\nu]^2 \right), 
\end{equation}
where $L_\mu \equiv (\partial_\mu U) U^\dagger$ and where
$R_\mu \equiv U^\dagger (\partial_\mu U)$ will be later required.
Finally, the Wess-Zumino-Witten (WZW) term 
${\cal L}_{\mbox{\scriptsize WZW}}$ is necessary 
to break the symmetry of Eq.(\ref{L0}) under $U\rightarrow U^{\dagger}$  
which is not a genuine symmetry of QCD. 
The corresponding action can be written locally as \cite{Witten83}
\begin{equation}
  \Gamma_{\mbox{\scriptsize WZW}}(U) = 
   C \int_{M^5} \mbox{Tr}( \hat{L}^5 )
   = C \int_{M^5} \mbox{Tr}( \hat{R}^5 )
\label{WZW0}\end{equation}
in a five-dimensional space $M^5$ whose boundary is ordinary 
space and time. The constant $C$ is determined as 
$C=-\frac{iN_c}{240\pi^2}$ so that the anomalous Lagrangian 
is consistent with the triangle anomaly for the process
$\pi^0\rightarrow \gamma\gamma$ when the number of colors $N_c=3$.
For compactness in the following we adopt the differential one-form notation 
with
$$
  \hat{L} = L_\mu d x^\mu = (dU) U^\dagger ,
$$
and
\begin{equation} 
  \hat{R} = U^\dagger dU = U^\dagger \hat{L} U.
\end{equation}

Gauging all but the WZW part of the Lagrangian enables 
the electroweak vector fields to be introduced via minimal coupling:
\begin{equation}
  D_\mu U = \partial_\mu U - i{\cal A}^L_\mu U + U (i{\cal A}^R_\mu),
  \label{covar_deriv}
\end{equation}
where 
\begin{eqnarray}
  {\cal A}^L_\mu 
  &=& 
  e Q A_\mu - \frac{g}{\sqrt2}(W_\mu^{+}\tau^{+} +W_\mu^{-}\tau^{-})
  \nonumber\\
  &&
  + \frac{g}{2\cos\theta_W}Z_\mu (\tau^3 - 2 Q \sin^2\theta_W),
  \\
  {\cal A}^R_\mu 
  &=&
  e Q A_\mu  - \frac{g}{2\cos\theta_W}Z_\mu (2Q \sin^2 \theta_W).
\end{eqnarray}
Here, $A_\mu$ is the electromagnetic field which couples with 
strength $Q=\mbox{diag}(2/3,-1/3)$, $W^{+,-}_\mu$ and $Z_\mu$ 
are the $SU(2)$ electroweak gauge fields, 
$g$ is the weak coupling constant, $\theta_W$ is the Weinberg 
angle, and $\tau^{\pm}=\frac12(\tau^{1} \pm i\tau^{2})$
with $\tau^a$ the Pauli matrices.

In contrast, as is well-known, minimal substitution
does not work in gauging the 
WZW action. The so-called trial and error Noether 
method \cite{Witten83} gives the gauged WZW action 
\begin{eqnarray}
  && 
  \tilde{\Gamma}_{\mbox{\scriptsize WZW}}(U,{\cal A}^L_\mu,{\cal A}^R_\mu) 
  \nonumber\\
  && \hskip 1em
  ={\Gamma}_{\mbox{\scriptsize WZW}}(U)
  + 5Ci \int_{M^4}\hskip -0.5em  \mbox{Tr} 
       (\hat{\cal A}^L \hat{L}^3 
       +\hat{\cal A}^R \hat{R}^3 )
  \nonumber\\
  && \hskip 2em
  - 5C \int_{M^4}\hskip -0.5em  \mbox{Tr} 
     \big(  (d\hat{\cal A}^L\hat{\cal A}^L 
           +\hat{\cal A}^Ld\hat{\cal A}^L)\hat{L}  
          \nonumber \\
        && \hspace{2cm}  + (d\hat{\cal A}^R\hat{\cal A}^R 
           +\hat{\cal A}^Rd\hat{\cal A}^R)\hat{R}
     \big)
  \nonumber\\
  && \hskip 2em
  + 5C \int_{M^4}\hskip -0.5em \mbox{Tr} 
     \big(d\hat{\cal A}^L dU \hat{\cal A}^R U^\dagger 
          -d\hat{\cal A}^R d(U^\dagger) \hat{\cal A}^L U 
     \big)
  \nonumber\\
  && \hskip 2em
  +\cdots.
\end{eqnarray}
Here we have shown only the terms relevant for the processes 
involving single neutral pion and one or two gauge bosons. 
A complete list of terms can be found in \cite{Witten83}.

The leptonic part of the Lagrangian is 
\begin{eqnarray}
    {\cal L}^{\mbox{\scriptsize lept.}}_{Z\nu\bar{\nu}} 
     &=& 
      - \frac{g}{2\cos\theta_W} Z_\mu 
        \bar{\nu}_e \gamma_\mu \frac{1-\gamma_5}{2} \nu_e 
      + \bar{\nu}_e (i\partial\hskip -0.5em\slash) \nu_e
   \nonumber\\
     && +\big( e \rightarrow \mu,\tau \mbox{ terms} \big).  
\end{eqnarray} 
Here again, we have presented only the relevant terms for 
the process of our concern, that is, $Z \rightarrow \nu\bar{\nu}$.
The neutrino field $\nu_e$ is constrained to the left-handed one; 
$\gamma_5 \nu_e^{} = - \nu_e^{}$. 

Finally, the Lagrangian should be complemented by the 
terms for the dynamics of the gauge bosons, namely
\begin{eqnarray}
  {\cal L}_{\gamma,Z,W^\pm} 
   &=& - \frac14\displaystyle \mbox{Tr} 
                 (F^{{\cal A}^L}_{\mu\nu}F^{{{\cal A}^L}\mu\nu}
                 +F^{{\cal A}^R}_{\mu\nu}F^{{{\cal A}^R}\mu\nu}) 
    \nonumber\\
    && - \frac12 \displaystyle\big( 
          M_W^2 (W^{+}_\mu W^{+\mu}+W^{-}_\mu W^{-\mu})
    \\
    && \hskip 2em  
          +M_Z^2 Z_\mu Z^\mu \big),
    \nonumber
\end{eqnarray}
where $F^{\cal L,R}_{\mu\nu}$ are field strength tensors and 
$M_W$ and $M_Z$ are the masses of the gauge bosons.

\section{$\pi^0 \rightarrow \nu\bar{\nu}$ in baryon-free space}
The lowest static energy configuration for baryon free space is simply 
a constant $U$. Without loss of a generality we can set $U=1$. 
Thus, the Lagrangian constructed in the previous section 
governs the dynamics of the pions in baryon free space
as it is. By expanding $U=\exp(i\vec{\tau}\cdot\vec{\pi}/f_\pi)$ up 
to a given order in the pion field we obtain the hadronic part 
of the lagrangian for the weak decay of the neutral pion as
\begin{eqnarray}
  {\cal L} 
  &=& \textstyle 
  \frac12 \partial_\mu \pi^a \partial^\mu \pi^a 
  -  \frac12 m_\pi^2 \pi^a\pi^a 
  \nonumber\\
  && 
  +\frac{f_\pi g}{2\cos\theta_W} Z_\mu \partial^\mu \pi^0 
  \label{pi0-vertex0}\\
  && 
  -\frac{N_c}{48\pi^2 f_\pi} \pi^0 \tan^2\theta_W \cos 2\theta_W g^2 
   \varepsilon^{\mu\nu\alpha\beta}
   \partial_\mu Z_\nu \partial_\alpha Z_\beta
  \nonumber\\ 
  && +\cdots.  
  \nonumber
\end{eqnarray} 
The Lagrangian yields proper vertices for $\pi^0 \rightarrow Z$ and 
$\pi^0 \rightarrow ZZ$, which contribute to the processes 
$\pi^0 \rightarrow \nu\bar{\nu}$ as shown in Fig.1.
Observe that there does not appear a term such as 
$\pi^0 \rightarrow W^{+} W^{-}$ despite fulfilling charge conservation. 
 
The amplitude for the process shown in Fig.1a is
\begin{equation}
{\cal M}_{\pi^0 \rightarrow Z \rightarrow \nu\bar{\nu}} 
 = \frac{G_F f_\pi}{\sqrt{2}}
   \bar{u}_\nu(p_1) p \hskip -0.5em \slash (1-\gamma_5) v_\nu(p_2),  
\end{equation}
where we have approximated the $Z$-boson propagator simply as 
$i/(p^2-M_Z^2) \sim -i/M_Z^2$ and $G_F=g^2/(8M_W^2)$.
$p_1=(\omega_1,\vec{p}_1)$ and $p_2 = (\omega_2,-\vec{p}_2)$
are the energy-momenta of outgoing neutrinos and $p=p_1+p_2$ is that 
of the incoming pion. 
For massless neutrinos
$ \bar{u}_\nu(p_1) \slash\hskip -0.5em  p_1  = 
\slash \hskip -0.5em p_2 v(p_2)=0$ so that the amplitude vanishes 
identically.

\begin{figure}
\begin{center}
\includegraphics[width=3cm,height=7cm,angle=270]{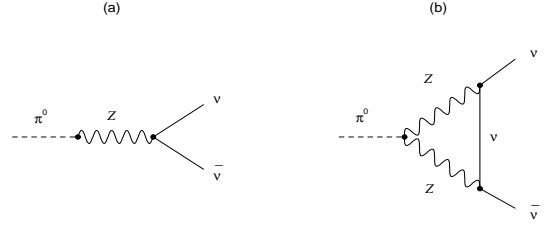}
\end{center}
\vskip -4ex
\caption{$\pi^0 \rightarrow \nu\bar{\nu}$ processes.}
\end{figure} 

The process of Fig.1b leads to
\begin{eqnarray}
{\cal M}_{\pi^0 \rightarrow ZZ \rightarrow \nu\bar{\nu}}
\!\!\!&=&\!\!\!
-i \frac{N_c g^2}{48\pi^2 f_{\pi}} \tan^2\theta_W \cos2\theta_W
\nonumber \\
&&\times  \varepsilon_{\mu \nu \alpha \beta} \int {{d^4k}\over{(2\pi)^4}}
{ k_1^{\alpha} k_2^{\beta} }
 \\
\!\!&&\!\! \times 
 { { \bar{u}(p_1) \gamma^{\mu} (1-\gamma_5) 
({k_2}\hskip -0.8em \slash -{p_2}\hskip -0.8em \slash) 
\gamma^{\nu} (1-\gamma_5) v(p_2) }  \over   
 {(k_1^2-M_Z^2) (k_2^2-M_Z^2) (k_2-p_2)^2} } 
\nonumber 
\end{eqnarray} 
where $k_1=k+p/2$, $k_2=-k+p/2$ represent the momenta
of the two internal Z-boson lines. The loop-momenta are
integrated only up to a cut-off $\Lambda\sim 1 {\rm{GeV}}$ 
for the low-energy effective theory. Taking the square
of the amplitude and tracing over the spinor structure gives
\begin{eqnarray}
|{\cal M}_{\pi^0 \rightarrow ZZ \rightarrow \nu\bar{\nu}}|^2
\!\!\!&=& \!\!\!{{N_c^2 g^4}\over {144\pi^4 f_{\pi}^2}}
\tan^4\theta_W \cos^2 2\theta_W
 \\
&& \!\!\!\times 
 I_{\Lambda}(p,p_1,p_2) 
(p^2 p_1 \cdot p_2 - 2 p\cdot p_1 p\cdot p_2)
\nonumber 
\end{eqnarray}
where $ I_{\Lambda}(p,p_1,p_2)$ comes from the loop-integration
but whose form is not important in view of the next result: 
in the pion rest frame $(p^2 p_1 \cdot p_2 - 2 p\cdot p_1 p\cdot p_2)$,
which arises from the spinor traces for massless neutrinos, 
vanishes identically, consistent with the helicity selection rule.

\section{$\pi^0 \rightarrow \nu\bar{\nu}$ in the dense skyrmion matter}
The Skyrme Lagrangian supports classical soliton solutions, skyrmions, 
whose topological winding number can be interpreted as a baryon 
number. 
With this conjecture, dense baryonic matter can be approximated 
as a system of skyrmions.
Let $U_0(\vec{r})=n_0+i\vec{\tau}\cdot\vec{n}$
parametrise the lowest energy configuration for a given baryon 
number density, where $n_0, \ {\vec n}$ are space-dependent
functions describing the ensemble of skyrmions. 
Then we can incorporate quantum fluctuations on top of this classical 
ground state configuration via the {\it ansatz} 
\begin{equation}
  U = \sqrt{U_\pi} U_0 \sqrt{U_\pi},
  \label{Ansatz}
\end{equation}
where $U_\pi = \exp(i\vec{\tau}\cdot\vec{\pi}/f_\pi)$. 
The trivial case of $U_0(\vec{r})=1$ corresponds to pions in 
baryon-free space. For the skyrmion 
matter encoded in $U_0(\vec{r})$ 
we take the crystal configurations studied in  
\cite{SkyrmionCrystal}. At low density, the skyrmion matter
is arranged in a face-centered-cubic crystal where a well-localised 
single skyrmion occupies each lattice site. 
As the density increases, the skyrmions begin to overlap. 
At higher density, the phase becomes the so-called ``half-skyrmion 
cubic crystal'', where about half of the baryon number of the original 
skyrmion leaks out to generate another well-defined dense object in the 
region of overlap. 
When the skyrmion matter is in an exact half-skyrmion phase, the spatial 
average value of the $U_0(\vec{r})$ vanishes identically.

Substituting the {\it ansatz} into the Lagrangian and 
expanding in fluctuations, we obtain a number of structures.
In the following we explicitly give only those terms
governing the dynamics of the fluctuations in the presence
of the background potentials which will be  
relevant to our discussion. Thus we obtain 
\begin{eqnarray}
  {\cal L} 
    &=& \textstyle 
    \frac12 G_{ab}(\vec{r}) \partial_\mu \pi^a \partial^\mu \pi^b 
    - \frac12 n_0 m_\pi^2 \pi^a \pi^a 
  \nonumber\\
    &&
    + \textstyle\frac12
    \varepsilon^{abc} \pi^a \partial_i \pi^b V_i^c(\vec{r}) 
  \nonumber\\
    && 
    + \frac{f_\pi g}{2\cos\theta_W} G_{33}(\vec{r}) 
      Z^\mu \partial_\mu \pi^0 
\label{effLag}  \\
    &&
    + \frac{N_c}{24\pi^2} \frac{1}{f_\pi} 
      \frac{g\sin^2\theta}{\cos\theta_W}
      \varepsilon^{\mu\nu\alpha\beta} Z_\mu \partial_\nu \pi^0 
      H_{\alpha\beta}(\vec{r}) 
  \nonumber\\
    && + \cdots, 
  \nonumber
\end{eqnarray}
where
\begin{eqnarray}
  G_{ab}(\vec{r}) 
    &=& 
    n_0^2 \delta_{ab} + n_a n_b,
  \\
  V_i^a(\vec{r}) 
    &=&
    \varepsilon^{abc} n_b \partial_i n_c
  \\
  H_{\alpha\beta}(\vec{r})
    &=&
    (n_0^2 + n_3^2) (\partial_\alpha n_1 \partial_\beta n_2
                    -\partial_\alpha n_2 \partial_\beta n_1) 
  \nonumber\\
    && - (n_2 \partial_\alpha n_1 - n_1 \partial_\alpha n_2)
         \partial_\beta (n_0^2 + n_3^2)
  \\
    &=& G_{33} \partial_\alpha V_\beta^3 
      + V_\beta^3 \partial_\alpha G_{33}.
  \nonumber
\end{eqnarray}
We stress that the terms not given here govern
the skyrmion dynamics summarised in these potentials,
and have been treated in earlier works.
In particular, formulae for
$n_0, {\vec n}$ are given in Eqs.(9-12) of \cite{LPMRV03} where
a different notation is used: 
$n_0\leftarrow {\bar \sigma}, n_i \leftarrow {\bar \pi}_i$.
 
We now consider whether it is possible for the process  
$\pi^0 \rightarrow \nu\bar{\nu}$ to occur 
through a one boson intermediate state, 
which is leading order in $G_F$.
To this end we need $G_{33}(\vec{r})$, which
comes from the kinetic term of 
the Lagrangian, and $H_{\alpha\beta}$ which originates in 
the WZW terms that are linear in ${\cal A}^L_\mu$ or ${\cal A}^R_\mu$. 
The contribution of $G_{33}$ is involved with the symmetric
(in $\mu\nu$) structure 
$g^{\mu \nu} Z_\mu \partial_\nu \pi^0$ while $H_{\alpha\beta}$ appears
with the antisymmetric (in $\mu\nu$)   
$\varepsilon^{\mu\nu\alpha\beta} Z_\mu \partial_\nu \pi^0$.

In a naive mean field approximation, we can replace the 
space-dependent potentials such as $G_{ab}(\vec{r})$ and 
$H_{\alpha\beta}(\vec{r})$ by their spatial averages
$\langle G_{ab}(\vec{r}) \rangle$ and 
$\langle H_{\alpha\beta}(\vec{r}) \rangle$; {\it viz.\/}
\begin{equation}
  \begin{array}{l}
     \langle G_{ab} \rangle 
       \equiv \displaystyle 
       \frac{1}{V} \int_V d^3x G_{ab}(\vec{r}) 
       = G \delta_{ab}, \\
     \langle V_i^a \rangle = 
     \langle H_{ab} \rangle = 0,
  \end{array}
\end{equation}
where $V$ is the volume of a unit cell of the skyrmion crystal 
over which the integration is carried out 
and $G$ is a constant. 
Due to the symmetric structure of the crystal solution 
$U_0(\vec{r})$, these averages of the potentials
$H_{\alpha\beta}(\vec{r})$ vanish.  
After renormalising the pion fields, $\pi^{*a} = \sqrt{G} \pi^a$, 
we obtain an {\em effective} Lagrangian as 
\begin{eqnarray}
   {\cal L}^* 
    &=& 
     \textstyle \frac12 \partial_\mu \pi^{*a} \partial_\mu \pi^{*a} 
     - \frac12 m_\pi^* \pi^{*a}\pi^{*a} 
   \nonumber\\
    && 
    +\frac{f^{*}_\pi g}{2\cos\theta_W} Z_\mu \partial^\mu \pi^{*0} 
   \label{pi0-vertex}\\
    &&
   +\cdots. 
   \nonumber    
\end{eqnarray}
Note that the structure of the Lagrangian is exactly the same as 
Eq.(\ref{pi0-vertex0}), that of the pions in baryon free space. 
Only the $\pi^0\rightarrow Z$ 
vertex strength is modified by a factor $f^*_\pi/f_\pi = \sqrt{G}$.
Lorentz symmetry is still conserved in this Lagrangian.
Thus, the $\pi^0\rightarrow Z$ vertex cannot contribute to 
$\pi^0 \rightarrow \nu\bar{\nu}$ as discussed in Sec. III. 

In \cite{LPRV04} is shown that Lorentz 
symmetry breakdown of pion dynamics in medium 
can be manifested only when higher order 
effects in the background potential beyond mean field 
approximation are incorporated. 
We now outline a perturbative expansion in
the background potentials.  
We decompose the Lagrangian into an unperturbed part, 
${\cal L}_{(0)}$ and an interaction part, ${\cal L}_I$.
We take ${\cal L}_0$ be the Lagrangian for free pion in baryon 
free space, namely 
\begin{equation}
  {\cal L}_0 = \textstyle \frac12 \partial_\mu \pi^a \partial^\mu \pi^a 
   +\frac12 m_\pi^2 \pi^a \pi^a, 
\end{equation}
and ${\cal L}_I$ the remaining terms in Eq.(\ref{effLag}).
The local potential $G_{ab}(\vec{r})$ in the kinetic 
term makes the quantization process somewhat nontrivial. Nevertheless the  
following steps are standard for perturbing in the 
spatial potentials - see \cite{Walecka}. 
The conjugate momenta of the pion fields $\pi^a$ are given by
\begin{equation}
  \Pi^a = \frac{\partial {\cal L}_0}{\partial \dot{\pi}^a} 
        = {\dot \pi}^a
\end{equation}
while the Hamiltonian is 
\begin{equation}
  {\cal H} = {\cal H}_0 + {\cal H}_I 
\end{equation}
where 
\begin{equation}
  {\cal H}_0 = \textstyle \frac12 
    (\Pi^a \Pi^a + \partial_i \pi^a \partial_i \pi^a )
     + \frac12 m_\pi^2 \pi^a \pi^a,
\end{equation}
and ${\cal H}_I = - {\cal L}_I$.

\begin{figure}
\begin{center}
\includegraphics[width=1.8cm,height=8cm,angle=270]{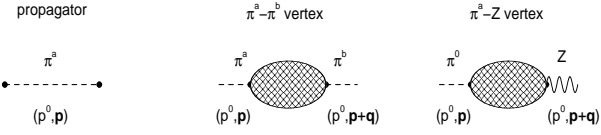}
\end{center}
\caption{Diagrammatic rules for the pion propagator and 
the interaction Hamiltonian.}
\end{figure}

This defines the pion propagator and the interaction Hamiltonian 
as schematically represented in Fig.2.  
The perturbation rule is now very simple:
\begin{enumerate}
\item Draw the diagrams for the corresponding physical process.
 The number of shaded blobs included in the diagram determines 
the order in perturbation.  
\item 
For each internal pion propagator, we have 
\begin{equation}
  \int d^4 p \ \frac{i}{(p^2 - m_\pi^2)}.
\end{equation}

\item The
$\pi^a(p)-\pi^b(p^\prime)$ vertex is given by
\begin{equation}
  \begin{array}{l}
     \delta^4 (p+\vec{q}-p^\prime) 
     \big\{ 
        -(p^2 - \vec{p}\cdot\vec{q})(G_{ab}(\vec{q})-\delta_{ab})
     \\
     \hskip 7em 
     +i \varepsilon^{abc} \vec{p}\cdot\vec{V}^c(\vec{q})
     \big\},
  \end{array}
\end{equation}
where $p$ and $p^\prime$ are the four-momenta of incoming and 
outgoing pions. We observe that the background medium is a source/sink for 
the additional momentum 
$\vec{q}$. 

\item The 
$\pi^0-Z$ vertex is
\begin{equation}
  \delta^4 (p+\vec{q}-p^\prime) 
  \left\{ 
     i p_\mu G_{33}(\vec{q})
    +i \varepsilon^{\mu\nu\alpha\beta} p_\mu H_{\alpha\beta}(\vec{q})
  \right\}
\end{equation}

\item 
As for the external fields, we follow standard 
conventions.

\end{enumerate}
The momentum components of the interactions are nothing but the 
Fourier expansion coefficients of the corresponding potentials, 
for example, 
defined as
\begin{equation}
  G_{ab}(\vec{q}) = \frac{1}{V} \int_V d^3 x 
            e^{i\vec{q}\cdot\vec{x}} G_{ab}(\vec{x}),
\label{fourierG}
\end{equation}
with similar definitions for $V_i^a(\vec{q})$ and 
$H_{\alpha\beta}(\vec{q})$ applying. Due to the crystal structure 
of the background field configuration, only discrete 
values of $\vec{q}$ are allowed,
\begin{equation}
\vec{q} = \frac{2\pi}L {\vec m}
\end{equation} 
where ${\vec m}$ label the spatial positions of the Skyrmions
in the FCC crystal and $L$ is the lattice spacing \cite{SkyrmionCrystal}. 
The naive mean field approximation corresponds
then to the $\vec{q}=0$ components of objects such as 
$G_{ab}(\vec{q})$.

\begin{figure}
\begin{center}
\includegraphics[width=1.8cm,height=8cm,angle=270]{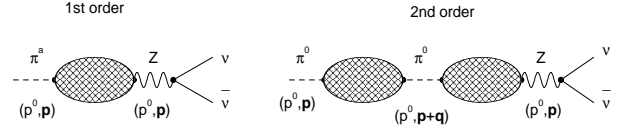}
\end{center}
\caption{
First and second order diagrams in perturbation in the
spatial potentials}.
\end{figure} 

As for the $\pi^0 \rightarrow \nu\bar{\nu}$ process, we can have 
the diagrams shown in Fig.3 up to second order in the 
background potential. 
Clearly the baryonic medium can impart energy-momentum
to the pion, which at the end of the day is why
the process $\pi^0\rightarrow \nu {\bar \nu}$ can proceed.
However the general problem with arbitrary momentum
imparted to the pion from the medium is difficult to
solve. We shall restrict ourselves to those contributions which
can be expressed in terms of a {\it local} $\pi^0\nu {\bar \nu}$
vertex, namely where the four-momentum of the incoming
pion is the equal to the total momentum of the outgoing
$\nu {\bar \nu}$. Certainly this is only a partial use
of the information in the formalism. We will see
that to leading order the effect will still be vanishing
but at second order in the potentials a non-zero
result can be obtained because of a reorganisation of
the momentum in the system between the two interactions
of the potentials.
 
In more detail then, the first order diagram reads
\begin{equation}
  {\cal M}^{(1)} 
     = \frac{G_F f_\pi}{\sqrt{2}} G_{33}(\vec{0})
       \bar{u}_\nu(p_1) p \hskip -0.5em \slash (1-\gamma_5) v_\nu(p_2),  
\end{equation}
which is the same as that for baryon free-space except for the factor 
$G_{33}(\vec{0})=\langle G_{33}\rangle =G$. 
At any rate, this amplitude vanishes identically 
as hinted above because 
$\bar{u}_\nu(p_1) p \hskip -0.5em \slash (1-\gamma_5) v_\nu(p_2)=0$   
for momenta $p=p_1+p_2$.  

The second order diagram yields 
\begin{eqnarray}
  {\cal M}^{(2)} 
     &=&  \displaystyle 
       \frac{G_F f_\pi}{\sqrt{2}}
       \sum_{\vec{q}}
       \frac{-(p^2 - \vec{q}\cdot\vec{p})
       G_{33}(\vec{q})}{(p+\vec{q})^2-m_\pi^2}
  \nonumber\\  
       &&
    \times   \bar{u}_\nu(p_1) \big\{ G_{33} (-\vec{q})(p \hskip -0.5em \slash 
       + \vec{q}\hskip -0.5em\slash)
       (1-\gamma_5)
  \nonumber\\
       && \hskip 1em
       +\varepsilon^{\mu\nu\alpha\beta} (p+\vec{q})_\mu 
       H_{\alpha\beta}(-\vec{q})
        \gamma_\nu
       (1-\gamma_5) \big\}  v_\nu(p_2)
  \nonumber\\
    & \approx& 
       \frac{G_F f_\pi}{\sqrt{2}} \delta 
       \bar{u}_\nu(p_1) \vec{p} \hskip -0.5em \slash 
       (1-\gamma_5) v_\nu(p_2), 
\end{eqnarray}
where 
\begin{equation}
  \delta= \textstyle \frac13 \displaystyle 
   \sum_{\vec{q}\neq 0}
   |G_{33}(\vec{q})|^2
\label{delta}
\end{equation}
In the last step we have used that 
(i) $\bar{u}_\nu(p_1) p \hskip -0.5em \slash (1-\gamma_5) v_\nu(p_2)=0$,    
(ii) $|G_{33}(\vec{q})|^2$ is even with respect to 
$\vec{q} \leftrightarrow -\vec{q}$, and  
(iii) only the first nonvanishing term 
in the expansion the pion momentum point 
$(p_0,\vec{p})=(m_\pi,\vec{0})$ need be kept.
Note that only the spatial component $\vec{p}$ appears in the 
equation, which indicates that the Lorentz symmetry is indeed broken. 
Also, the contributions from $H_{\alpha \beta}$ of Eq.(\ref{effLag})
only appear at higher order in this pion momentum expansion.   

Thus, up to the second order in the background potential, 
we have the nonvanishing result
\begin{equation}
  \sum_{\mbox{\scriptsize spin}}  |{\cal M}^{(1)+(2)}|^2 
   = 4 G_F^2 |\chi_{\rm{Sk}}|^2 \omega_\pi^2 
     (\omega_1 \omega_2 + \vec{p}_1\cdot\vec{p}_2),
\label{amplsquared}
\end{equation} 
with $\chi_{\rm{Sk}}=f_\pi \delta + {\cal O}(m_{\pi}^2)$,
$\omega_\pi=\sqrt{m_\pi^2+v_\pi^2 p^2}$ the pion energy
and $\omega_{1,2}$ the corresponding energy for the
neutrino and anti-neutrino, the sum of which equals that for
the pion.
This is the result for one neutrino flavour; for more flavours the
result is multiplied by $N_{\nu}$.

Eqs.(\ref{delta},\ref{amplsquared}) are the main formal
results of this work. We plot numerical results
for the density dependence of $\delta$ in Fig.4.
We now discuss this with
respect to other work in the literature.

\begin{figure}
\begin{center}
\includegraphics[width=4cm,height=6cm,angle=270]{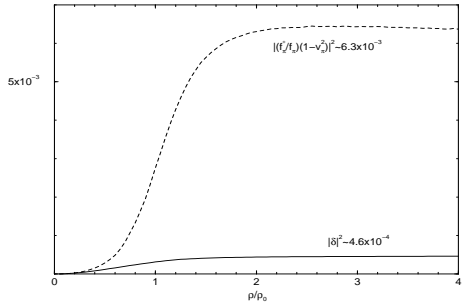}
\end{center}
\vskip -4ex
\caption{Density dependence of $\delta$ (solid line) and the 
corresponding quantity involving the pion velocity from
\cite{LPRV04}. Here $\rho_0$ is the density of normal nuclear matter. }
\end{figure}

To begin with Eq.(\ref{amplsquared}) 
is formally the same as the corresponding quantity 
in \cite{JPS02} but where for us $\chi_{\rm{Sk}}$ depends
on the medium via $\delta$ and on the {\it vacuum} value
for the decay constant $f_{\pi}$. However in \cite{JPS02}
$\chi_{\rm{Sk}}$ is replaced by a quantity which we shall designate 
$\chi_{JPS}$ itself depending on the velocity in medium
of the CFL generalised pion, $v_{\pi}$.
This velocity in \cite{JPS02}
measures in turn the breakdown of Lorentz symmetry
as reflected in the non-equivalence of spatial $(f_S)$ and temporal
$(f_T)$ components of the corresponding pion decay constant in medium, 
\begin{equation}
  |\chi_{JPS}| = |f_T - f_S| = |f_T(1-v_\pi^2)|.
\end{equation}
Thus despite studying quite different phases (hadronic vs CFL)
we find formal agreement to second order in the potentials,
up to what appears as a discrepancy: 
the absence of a medium generated distinction between
components of the decay constant. However in a study of
the pion propagator in dense skyrmionic matter in \cite{LPRV04}  
it was shown by resummation of the corresponding expansion in potentials
to infinite order that 
the in-medium pion velocity is given by
\begin{equation}
   1-v_\pi^2 = \sum_{\vec{q}\neq 0} 
   \frac{\displaystyle \sum_{a=1,2} |V_a^3(\vec{q})|^2 
   + \textstyle \frac13 \vec{q}^2 
     \displaystyle \sum_{a=1,2,3} |G_{3a}(\vec{q})|^2}{q^2}.
\label{pionvel}
\end{equation}
Comparing this to the present result for $\delta$, Eq.(\ref{delta}), one 
expects that a similar, though technically more difficult,
resummation in the context of $\pi^0 \rightarrow \nu\bar{\nu}$
will reveal the dependence of this decay in medium
on the pion velocity and thus in turn on the inequivalence
of $f_T$ and $f_S$ in medium. 
It is important to stress that the isospin mixing terms in
the numerator of Eq.(\ref{pionvel}) cannot appear in our
present calculation of Eq.(\ref{delta}) to second order
in potentials, and can only occur after resummation.  
The analogous combination
of in-medium quantities, $f^*_{\pi}$ and $v_{\pi}$
as obtained in \cite{LPRV04} are indicated with the dashed
curve in Fig.4 for comparison. 

We see from Fig.4 our present result $|\delta|^2\sim 4.6\times 10^{-4}$
representing a significant density enhancement in the hadronic phase 
in the decay rate of $\pi^0\rightarrow \nu {\bar \nu} $ as compared to 
$(m_{\nu}/m_{\pi})^3 \sim 10^{-9} $ for keV neutrinos.
Resummation effects, were they to mirror the results of
\cite{LPRV04}, would amplify this further.   
In \cite{LPRV04}, the pion velocity is obtained in the 
range of $0.8 < v_\pi \leq 1$. This leads to
a larger $1-v_{\pi}^2$ than the $|\delta|$ obtained
in our second order calculation, but is still smaller than
that for the CFL phase in \cite{JPS02} where  
$v_\pi^2=1/3$. 
 
\section{Conclusion}
We have studied the neutral pion decay into a
neutrino-antineutrino pair. In baryon free space, the process 
is forbidden by helicity conservation for massless neutrinos.  
Though new vertices $\pi^0 \rightarrow Z$ and $\pi^0 \rightarrow ZZ$ emerge 
when the Skyrme Lagrangian is gauged, they also lead to a null result but
for the square of the amplitude. 
On the other hand, we have shown that Lorentz symmetry
breaks down due to the absolute frame of a background dense medium enabling 
$\pi^0 \rightarrow \nu\bar{\nu}$ 
which plays a crucial role in astrophysical phenomena.  

Our result formally matches with \cite{JPS02}, which
however dealt with the colour-flavour-locked phase
of matter at very high densities.
Thus our work shows that such a pion pole mechanism already can be
manifest even in the lower density hadronic phase.

Our result is the first study of the
$\pi^0 \rightarrow \nu\bar{\nu}$ in Skyrme models,
which elegantly combine meson and baryon
dynamics. The limitations of the Skyrme
model, in particular the provisional use of the 
unrealistic Skyrme crystal, 
mean our result is still qualitative.
We stress though that the discrete translational
symmetry of the crystal plays no role in these results,
only its general isotropy which is anyway realistic.
The mechanism for the decay is precisely the same as  
that hypothesised for the CFL phase and 
already can be exhibited for massless neutrinos in
the hadronic phase. To the order in perturbation
in background potentials studied here and  
retaining only those contributions to an
effective local $\pi^0\nu{\bar\nu}$ vertex, the result formally
agrees with that in the CFL phase.  
Moreover, the mechanism provides for significant
enhancement of the decay above the contribution
that a keV order neutrino mass would provide.
Its role in the overall process 
$\gamma\gamma\rightarrow \pi^0 \rightarrow \nu\bar{\nu}$ 
is not yet clear: in \cite{KP04} we have observed a
suppression in $\Gamma(\gamma\gamma\rightarrow \pi^0)$
while we have yet to consider the pion propagator
modifications in the spirit of \cite{ANV03}. 
Moreover introduction finite temperature effects
is necessary to enable a more detailed application to 
neutrino emissivity in compact stars.

\section*{Acknowledgements}
B-Y.P. is grateful for the hospitality of the Special Research Centre for
the Subatomic Structure of Matter (CSSM) at the University of Adelaide.
J.D.C. thanks the CSSM for support and Tony Williams 
for informative discussions. A.C.K. is supported by
the Australian Research Council.

\end{document}